    \documentclass[12pt,psf,epsf]{article}
\usepackage[centertags]{amsmath}
\usepackage{amssymb}
\usepackage{graphicx}
\usepackage{epsfig}
\usepackage{ulem}
\usepackage[english]{babel}
\usepackage{array}
\usepackage{amsthm}
\usepackage{latexsym}
\usepackage[mathcal]{euscript}
\pdfoutput=1
\usepackage{epsfig}

 \usepackage{jheppub}
 \usepackage{hyperref}

\usepackage{subfigure}
\usepackage{psfrag}

\usepackage{epsfig}
\usepackage[latin1]{inputenc}
\usepackage{float}
\usepackage{graphicx}
\usepackage{cancel}
\usepackage{mathrsfs}
\usepackage{amssymb}
\usepackage{amsfonts}
\usepackage{amsmath}
\usepackage{slashed}
\usepackage{graphicx}
\usepackage{bm}
\usepackage{color}
\newcommand{\be}{\begin{equation}}
\newcommand{\ee}{\end{equation}}
\newcommand{\bea}{\begin{eqnarray}}
\newcommand{\eea}{\end{eqnarray}}
\newcommand{\bwt}{\begin{widetext}}
\newcommand{\ewt}{\end{widetext}}

\newcommand{\nn}{\nonumber}

\newcommand{\bi}{\begin{itemize}}
\newcommand{\ei}{\end{itemize}}

\usepackage{setspace}

\definecolor{dgreen}{rgb}{0.,0.6,0.}

\begin{document}

\title {Butterfly velocity and chaos suppression in de Sitter space}

\author{Dmitry S. Ageev}
\affiliation{Steklov Mathematical Institute, Russian Academy of Sciences, Gubkin str. 8, 119991
Moscow, Russia}
\emailAdd{ageev@mi-ras.ru}

\abstract{ In this note we study the holographic CFT in the de Sitter static patch at finite temperature $T$ and chemical potential. We  find that butterfly velocity $v_B$ in such field theory degenerates for all values of the Hubble parameter $H$ and $T$.  We interpret this as a chaos disruption caused by the interplay between the expansion of chaotic correlations constrained by $v_B$ and effects caused by de Sitter curvature. The chemical potential restores healthy butterfly velocity for some range of temperatures. Also, we provide some analogy of this chaos suppression with the Schwinger effect in de Sitter and black hole formation from shock wave collision. 

}
\maketitle

\newpage

\section{Introduction} \label{sec:intro}
Chaos, and especially quantum chaos, has always been an intriguing topic in theoretical physics, which attracted attention since the early ages of quantum mechanics \cite{Einstein,gutz}. 
Recently, a criterion to characterize the chaotic properties of the quantum system has been proposed \cite{larkin,Shenker:2013pqa,Roberts:2014isa, Shenker:2013yza,Maldacena:2015waa,Shenker:2014cwa}. It is based on the notion of so-called out-of-time ordered
correlator (OTOC) and related to the exponential growth of squared commutators of  operators inserted at different spacetime points
\be 
\left\langle\left[{\cal O} \left(t,x\right), {\cal O}(0,0)\right]^{2}\right\rangle_{\beta} \sim   e^{\lambda_{L}\left(t-t_{*}-|x-y| / v_{B}\right)},
\ee 
where $t_*$ is  the scrambling time, $\lambda_L$ is the Lyapunov exponent and $v_B$ is the butterfly velocity constraining spatial chaos spreading inside the effective lightcone. Also, in \cite{Shenker:2013pqa,Roberts:2014isa, Shenker:2013yza, Maldacena:2015waa, Shenker:2014cwa}, it was argued that  quantum chaos has an intimate relation to quantum gravity and the triple  $t_*, \lambda_L,v_B$  is calculable from the shock wave gravity solutions in dual theory. In many cases, the butterfly velocity dependence on temperature has a simple polynomial form. For example, for quite general hyperscaling geometry, it has the form  $v_B\sim T^{1-1/z}$, where $z$ is  Lifshitz  exponent \cite{Blake:2016wvh,Roberts:2016wdl}. For a review of different aspects of recent developments in the quantum chaos and  holography see \cite{Jahnke:2018off}.

In \cite{Ahn:2019rnq}, it was shown that the dependence of butterfly velocity of holographic CFT on hyperbolic space at finite temperature   is more involved, indicating  that the curved space is an interesting testing ground for holographic quantum chaos calculational methods\footnote{See \cite{Ahn:2020bks} for the scalar and vector field pole-skipping analysis.}. While the hyperbolic spaces have been naturally associated with chaotic behavior for a long time before holography \cite{anosov}, their ``anti-chaotic'' positively curved counterparts are relatively unexplored \cite{gutz}. As a prototypical example of a positively curved background, one can consider the de Sitter space, which is especially important in the investigations of quantum gravity and cosmology  \cite{Hartle:1983ai,Witten:2001kn,Strominger:2001pn}. Since the de Sitter space is positively curved, one might naively expect that the chaos will disappear for some particular temperature of CFT or the Hubble parameter. The example of transition away from the chaotic behavior has been observed in  \cite{Ageev:2018msv,Ageev:2018tpd}  for the suppression of charged operator spreading in the case of a system at finite chemical potential.  

$\,$

In this note, we study the behavior of the butterfly velocity $v_B$ in a CFT at finite temperature in the de Sitter space with the Hubble parameter $H$ using holographic correspondence. As a dual background, we take the ``elliptic'' black hole - the black hole defined by de Sitter metric placed on the asymptotic boundary   (see \cite{Emparan:1998he,Birmingham:1998nr,Cai:1998vy,Chamblin:1999tk} as an example and references therein). Previous works on holography  with the de Sitter metric on asymptotic boundary \cite{Marolf:2010tg} addressed different aspects of duality including holographic Schwinger effect \cite{Fischler:2014ama}, entanglement entropy \cite{Fischler:2013fba}, Brownian motion \cite{Fischler:2014tka} etc. In \cite{Ahn:2019rnq} the chaotic properties for a theory dual to hyperbolic black hole have been considered. By analogy with the asymptotic hyperbolic boundary considered in \cite{Ahn:2019rnq}, the butterfly velocity calculation for our case  effectively reduces to the study  of massive scalar field solutions in de Sitter space with  the mass defined by the black hole horizon location. As it is known, the massive scalar field in de Sitter has the transition from oscillating to decaying behavior for some certain field mass and the Hubble parameter. The same mechanism  effectively drives the transition to the imaginary butterfly velocity here and which we interpret as the disruption of chaotic properties. Also we provide the derivation of the butterfly velocity based on the pole-skipping analysis. It turns out that positive curvature leads to the imaginary butterfly velocity for all values of temperature. However, the inclusion of the chemical potential restores the chaotic properties (i.e. real-valued butterfly velocity) for some restricted range of parameters. 

This note is organized as follows. In Sec.\ref{sec:ds-1} we introduce the gravitational setup and describe the effect of chaos disappearance for strongly coupled CFT at finite temperature in de Sitter space. In Sec.\ref{sec:disc} we discuss this result and demonstrate some parallels with other effects that take place  in de Sitter. 
\newpage
\section{Quantum chaos suppression in de Sitter space}\label{sec:ds-1}
\subsection*{Topological black hole and de Sitter on the boundary}
We consider  $(d+1)-$dimensional Einstein-Hilbert action
\be \label{eq:action}
S=\frac{1}{16 \pi G_{N}} \int d^{d+1} x \sqrt{-g}\left(R+\frac{d(d-1)}{L^{2}}\right),
\ee 
and the solutions of this theory of the form
\be \label{eq:metr} 
d s^{2}=-f(r) d t^{2}+\frac{d r^{2}}{f(r)}+r^{2} d \Sigma_{d-1}^{2},
\ee 
where  $L$ is the  AdS length scale and $\Sigma$ is some fixed curved spacetime.  Our main focus will be on the case when  $\Sigma$ is  $(d-1)$--dimensional (Euclidean) de Sitter static patch with the metric
\be \label{eq:metrDS}
d\Sigma_{d-1}^2={\left(1-\frac{\chi^2}{\ell^2}\right)}d\tau^2+\frac{d\chi^2}{1-\chi^2/\ell^2}+\chi^2d \Omega_{d-2}^{2},
\ee 
and the temperature $T_{dS}$ and the Hubble parameter $H$ 
\be
T_{dS}=\frac{1}{2\pi\ell}=\frac{H}{2\pi},\,\,\,\,\,\,\,\,H=\frac{1}{\ell}.
\ee 
Notice, that de Sitter is non-dynamical in our setting and is fixed as asymptotic boundary of topological black hole \eqref{eq:metr}. This is different from dS/CFT correspondence where de Sitter is dynamical and the hypotetical CFT lives on the future boundary of de Sitter \cite{Strominger:2001pn}.
For this choice of $d\Sigma_{d-1}$ the function $f(r)$  which solves \eqref{eq:action} has the form
\be 
f(r)=\frac{1}{\ell^2}+\frac{r^2}{L^2}-\frac{r_0^{(d-2)}}{r^{d-2}}\left(\frac{1}{\ell^2}+\frac{r_0^2}{L^2}\right),
\ee 
where   $r_0$ is the horizon location (see, for example, \cite{Birmingham:1998nr} and references therein for a discussion of this type of solutions). 

$\,$

From the viewpoint of holographic correspondence, the metric \eqref{eq:metr} is dual to a strongly coupled CFT at finite temperature defined on $\Sigma \times {\mathbb{ R}}$, where $\Sigma$ is given by metric \eqref{eq:metrDS}.  The temperature of black hole \eqref{eq:metr} is related to the horizon location $r_0$ as
\be \label{eq:temp}
T=\frac{f^{\prime}(r_0)}{4\pi}=\frac{1}{4\pi}\left( \frac{d-2}{l^2 }\frac{1}{r_0}+\frac{d }{L^2} r_0\right),
\ee 
and  the temperature dependence on $r_0$ is non-monotonous with the minimum at $r_m$
\be 
r_m=\frac{L}{l} \sqrt{\frac{d-2}{d}},\,\,\,\, T_m=\frac{\sqrt{d(d-2)} }{2 \pi  l L},
\ee 
where $T_m$ is the corresponding minimal temperature. In what follows we denote $\beta$ the inverse temperature of boundary $CFT$ and $\beta_{dS}$ the inverse temperature of de Sitter.

$\,$

 The butterfly effect on the gravity side of the holographic correspondence is related to the fact that the energy of an infalling particle has exponential behavior at late times \cite{Shenker:2013pqa,Roberts:2014isa,Shenker:2013yza}, and the backreaction of this particle creates a shock wave leading to the growth of certain operator commutators. The Lyapunov exponent and butterfly velocity calculation can be performed by consideration of the appropriate shock wave background. In \cite{Grozdanov:2017ajz,Blake:2018leo} it was noticed that the decoupling of the equation on shock wave profile (function $h$ below) and the decoupling of certain infalling gravitational perturbation in the near-horizon region has a similar origin, thus providing us access to the Lyapunov exponent and butterfly velocity. Their values are encoded in the special points of energy density correlators and can be found by the near-horizon analysis of gravitational perturbations. This method is  often called a pole-skipping analysis.  We present calculations by both of these methods for dual of \eqref{eq:metr} in the rest of this section.
 \subsection*{Shock wave calculation}
 To perform the shock wave calculation we   rewrite the metric \eqref{eq:metr} in the Kruskal coordinates 
 \be \nn
 d s^{2}=2 A(U V) d U d V+r^{2}(U V) d \Sigma_{d-1}^{2},\,\,\,\,A(U V)=\frac{\beta^{2}}{8 \pi^{2}} \frac{f(r(U V))}{U V}.
 \ee 
Following \cite{Shenker:2013pqa, Roberts:2014isa} and  \cite{Ahn:2019rnq} one can show that  the  effect of the perturbation in the form of a localized shock wave  can be taken into account by the shift in $V$ coordinate
\be \nn
d s^{2} \rightarrow d s^{2}+h_{V V} d V^{2},\,\,\,\,\,\,h_{V V}=\frac{16\pi G_N A_0}{r_0^{d-1}}  \delta(V) h\left(d\left(\mathbf{x}, \mathbf{x}^{\prime}\right)\right),
\ee 
where $A_0=A(0)$ and the explicit form of $h_{VV}$ is to be defined below\footnote{We refer reader to \cite{Ahn:2019rnq} for all details concerning the shock wave solution analysis and derivation of butterfly velocity for hyperbolic black holes. Also, it is worth to notice that in \cite{Ahn:2019rnq} authors considered fixed $AdS$ scale $\ell=1$.}. It is important to stress that $h$ is the function of geodesic distance $d\left(\mathbf{x}, \mathbf{x}^{\prime}\right)$ on the manifold $\Sigma$ (in our  case    given by \eqref{eq:metrDS}). The case when $\Sigma$ is the AdS space has been described  in \cite{Ahn:2019rnq}, and it was shown that $h$ is  defined by the solution of the wave equation  on $\Sigma$  
\be
\left[\square_{\Sigma_{d-1}}-m_{scr}^2\right] h\left(d\left(\mathbf{x}, \mathbf{x}^{\prime}\right)\right)=0,\,\,\,\,\,\,\,\, m_{scr}^2=\frac{2\pi r_0}{\beta}(d-1),
\ee 
where the shock wave profile behaves as
\be 
h(\chi)=c_{1} e^{-\mu d\left(\mathbf{x}, \mathbf{x}^{\prime}\right)},
\ee 
for large values of $d\left(\mathbf{x}, \mathbf{x}^{\prime}\right)$. The AdS space on the asymptotic boundary  results in $\mu=\mu_{Ad S}=\frac{1}{2}\left(d-2+\sqrt{(d-2)^{2}+4 m_{s c r}^{2} \ell^{2}}\right)$ with $v_B=2\pi T/\mu_{AdS}$. Here following \cite{Blake:2018leo} we call $m_{scr}$ a screening length. From this result it is straightforward to obtain the butterfly velocity for CFT living in de Sitter on asymptotic boundary 
\be \label{eq:vbmuds}
v_B=\frac{2\pi}{\beta \mu_{dS}},
\ee 
where $\mu_{dS}$ is expressed in terms of $\ell$, $L$ and $r_0$ as a ``conformal dimension'' of the operator with ``mass'' $m_{scr}$
\be \label{eq:muds}
\mu_{dS}=\frac{1}{2}\left(d-2+\sqrt{(d-2)^2-4m_{scr}^2\ell^2}\right).
\ee 
 The critical horizon value which is defined as
 \be \label{eq:crit}
 (d-2)^2-4m_{scr}^2\ell^2=0,
 \ee 
  separates the real-valued and imaginary butterfly velocities. The solution of this equation is given by
 \be
r_{cr}^2=\frac{(2-d) L^2}{2 (d-1) l^2},
\ee 
which shows that \eqref{eq:crit} has no real roots for $d>2$. This means that at finite temperature the buttefly velocity in de Sitter space value is complex-valued. 
 \subsection*{Pole-skipping calculation} 
 Now derive the butterfly velocity by pole-skipping analysis. The holographic pole-skipping analysis is a powerful method to calculate the special points in momentum space $\omega_*, k_*$ defined as follows. For a general retarded Green function of the form
\be 
G^{R}(\omega, k)=\frac{b(\omega, k)}{a(\omega, k)},
\ee
where zeroes of $a(\omega,k)$ corresponds to the dispersion relation $\omega=F(k)$. 

Points $\omega_*, k_*$  are defined as \begin{equation}
\begin{aligned}
&a\left(\omega_{*}, k_{*}\right)=0 \\
&b\left(\omega_{*}, k_{*}\right)=0.
\end{aligned}
\end{equation}
 The ``pole-skipping'' means, that dispersion relation defines a line of poles, which are ``skipped'' at special point $\omega_*, k_*$, because they are zeroes of $a$ and $b$ simultaneously. In holography, the pole-skipping points of two-point functions leave an imprint on the gravity side as the infalling plane waves with the special frequency and momentum \cite{Blake:2018leo}. Using pole-skipping points related to stress-energy correlators one can extract the Lyapunov exponent and butterfly velocity  \cite{Grozdanov:2017ajz,Blake:2018leo}. They are expressed in terms of $\omega_*$ and $k_*$ as
 \be 
 \omega_{*}=i \lambda_{L}, \quad k_{*}=i \frac{\lambda_{L}}{v_{B}}.
 \ee 
 In the case of asymptotically flat space boundary, the spatial part of the perturbation is a plane wave.   The key difference with the asymptotically flat boundary is that instead of plane waves one has to consider another ansatz for perturbations.   This was noticed for negatively curved boundary in \cite{Ahn:2019rnq} and our case closely follow their derivation.
  
  $\,$
  
We consider  the infalling gravitational perturbations
 \be 
 \delta g_{\mu \nu}(v, r, \tau, \chi,\Omega_{d-3})=\delta \bar{g}_{\mu \nu}(r, \tau, \chi,\Omega_{d-3}) e^{-i \omega v},
 \ee 
in the  near-horizon region of metric \eqref{eq:metr} rewritten in ingoing Eddington-Finkelstein  coordinates
\be 
ds^2=-f(r)dv^2+2dv dr+r^2d\Sigma_{n-1}^2.
\ee 
One can check that for a special frequency $\omega=\omega_*=2\pi T i$   the equation of motion for $g_{vv}^{(0)}$ component  corresponding to the near-horizon region $r\rightarrow r_0$
\begin{gather}
\delta \bar{g}_{vv}(r,  \tau, \chi,\Omega_{d-3})=\\=\delta g_{vv}^{(0)}(\tau,\chi,\Omega_{d-3})+\delta g_{vv}^{(1)}(\tau,\chi, \Omega_{d-3})\left(r-r_{0}\right)+...
\end{gather} 
decouples from the other components and leads us to a single equation of motion
\be \label{eq:deg}
m_{scr}^2\cdot\delta g_{vv}^{(0)}(\tau,\chi,\Omega_{d-3}) -\square_{dS_{d-1}}\cdot \delta g_{vv}^{(0)}(\tau,\chi,\Omega_{d-3})=0.
\ee 
As was noticed before we fix the special form of perturbation which is different from the usual plane wave
 \be 
 g_{vv}^{(0)}(\tau,\chi,\Omega_{d-3})=Y_{k}^{(d-2)}\left(-H \chi, \tau, \Omega_{d-3}\right),
 \ee 
 where we denote $Y_{k}^{(d-2)}\left(-H \chi,\tau, \Omega_{d-3}\right)$  the generalized spherical function \cite{Sfetsos:1994xa}. Notice, that here we use the metric defined in the form \eqref{eq:metrDS}, so for example for $d=3$ we have explicit form of  $Y_{k}^{(1)}(\chi,\phi)\sim P_k^M\left(-H \chi\right)\cdot e^{i\tau H M}$, where $P_k^M$ is an associated Legendre function. 
 
 Using the property that it is the  eigenfunction with respect to $\square_{dS_{d-1}}$ with eigenvalues defined by
 \begin{gather}
 \square_{dS_{d-1}} Y_{k}^{(d-2)}\left(- H \chi, \tau, \Omega_{d-3}\right)=\\=-H^2 k(k+d-2) Y_{k}^{(d-2)}\left(-H \chi, \tau,\Omega_{d-3}\right),
 \end{gather}
 we obtain the condition on $k$ when \eqref{eq:deg} vanishes identically
 \be  \label{eq:Lfix}
 k_*(k_*+d-2)+\frac{1}{H^2} m_{scr}^2=0,
 \ee 
 leading to sought pole-skipping point $k=k_*$.
 
 Using identity $H=1/\ell$ one can see that the explicit solution of \eqref{eq:Lfix} of the form
 \be \label{eq:Lds}
k_*=-\frac{1}{2}\left(d-2+\sqrt{(d-2)^2-4m_{scr}^2\ell^2}\right),
\ee 
  coincides with  $\mu_{dS}$ defined by $\eqref{eq:muds}$ up to a sign.  By analogy with  \cite{Ahn:2019rnq} again, we see that butterfly velocity is defined by
  \be 
  v_B=-\frac{2\pi T}{k_*}.
  \ee 
\subsection*{Turning on chemical potential}
Now turn on the chemical potential in dual theory which corresponds to  the charged topological black hole \cite{Cai:1998vy,Chamblin:1999tk} in the bulk. Einstein-Maxwell action has the form 
\be \label{eq:actionch}
S=\frac{1}{16 \pi G_{N}} \int d^{d+1} x \sqrt{-g}\left(R+\frac{d(d-1)}{L^{2}}+F^2\right),
\ee  
where $F$ is the Maxwell field strength.
The charged topological black holes solution  has the metric \eqref{eq:metr} with another emblackening function 
\be  \label{eq:fchar}
f(r)=\frac{1}{l^2}+\frac{r^2}{L^2}-\left(\frac{r_0}{r}\right)^{d-2} \left(\frac{q^2}{ r_0^{2d-4}}+\frac{1}{l^2}+\frac{r_0^2}{L^2}\right)+\frac{q^2}{ r^{2d-4}},
\ee 
and with the pure  electric gauge potential given by
\be
A_t=\left(-\frac{1}{c} \frac{q}{r^{d-2}}+\frac{1}{c} \frac{q}{r_{0}^{d-2}}\right),\,\,\,\,c=\sqrt{\frac{2(d-2)}{d-1}}.
\ee
The temperature now is given by
\be 
T=\frac{1}{4 \pi}\left(\frac{d-2}{l^2 r_0}+\frac{d r_0}{L^2}-(d-2) q^2 r_0^{3-2
   d} \right),
\ee 
and  condition $T>0$ defines the critical charge
\be 
q_{cr}^2=r_0^{2 d-4} \left(\frac{d r_0^2}{(d-2)
   L^2}+\frac{1}{l^2}\right).
\ee 
In contrast with $q=0$ now temperature has no minimal value and extends to zero for some $q=q_{cr}$. For $q<q_{cr}$ the spacetime describes charged black holes, while $q>q_{cr}$ corresponds to a naked singularity. The formulae for butterfly velocity \eqref{eq:vbmuds} and \eqref{eq:muds} work as well for $f(r)$ defined by \eqref{eq:fchar} with the corresponding screening length $m_{scr}$. 

\begin{figure}[h!]
\centering
\includegraphics[height=5.5 cm]{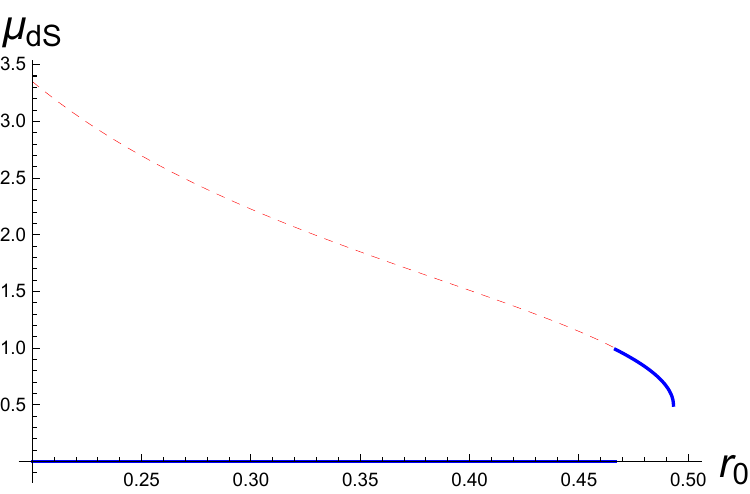}
\includegraphics[width=6 cm]{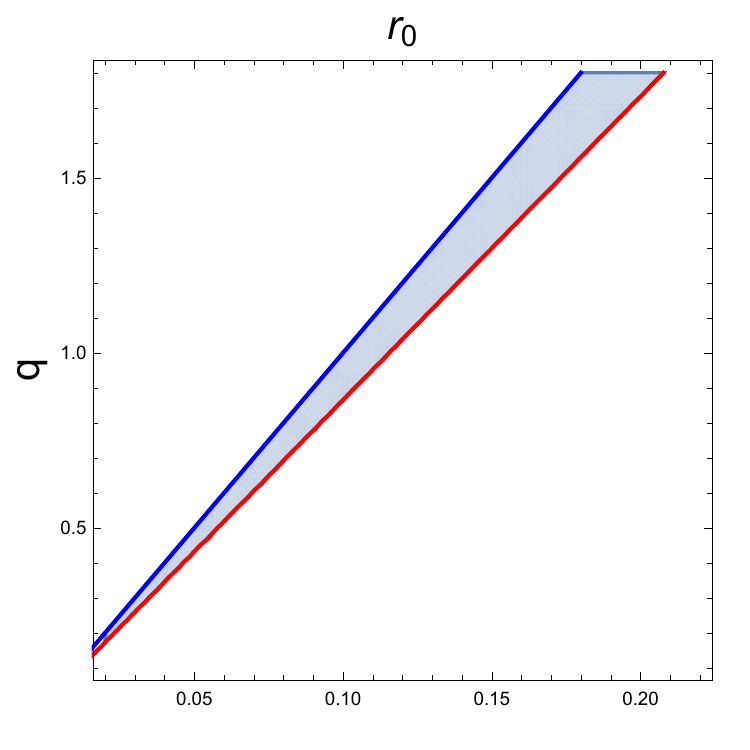}
 \label{fig:diagvb}
  \caption{Left plot: the dependence of $\mu_{dS}$ on $r_0$. Blue curve indicates   real-valued $\mu_{dS}$ which satisfy $q<q_{cr}(r_0)$ leading to healthy butterfly velocity, while the red dashed curve describes region where $q>q_{cr}(r_0)$ with $q=0.6$.
    Right plot: Blue curve is the critical charge $q=q_{cr}$ dependence on $r_0$, red curve is given by $q^2={\cal Q}$, shaded region corresponds to healthy butterfly velocity. For both plots $\ell=L=1$.}
\end{figure}

The charge value $q^2={\cal Q}$ defined by condition $(d-2)^2=4m_{scr}^2\ell^2$ separates imaginary and real-valued butterfly velocities and is given by equation
\be
{\cal Q}=\frac{1}{2} d r_0^{2 d-4} \left(\frac{1}{(d-1) l^2}+\frac{2
   r_0^2}{(d-2) L^2}\right)
\ee 
which is real-valued giving rise to a  real-valued buttefly velocity for some charge values.
In Fig.\ref{fig:diagvb}  we present curves $q=q_{crit}$ and $q^2={\cal Q}$ and one can see that there is a narrow  region of parameters $q$ and $r_0$ which defines real-valued butterfly velocity and satisfies the constraint $q<q_{cr}$.
This region is situated closely to the critical charge, which means  that  large chemical potential restores chaotic behavior for cold enough CFT in de Sitter.  
\begin{figure}[h!]
\centering
\includegraphics[height=5  cm]{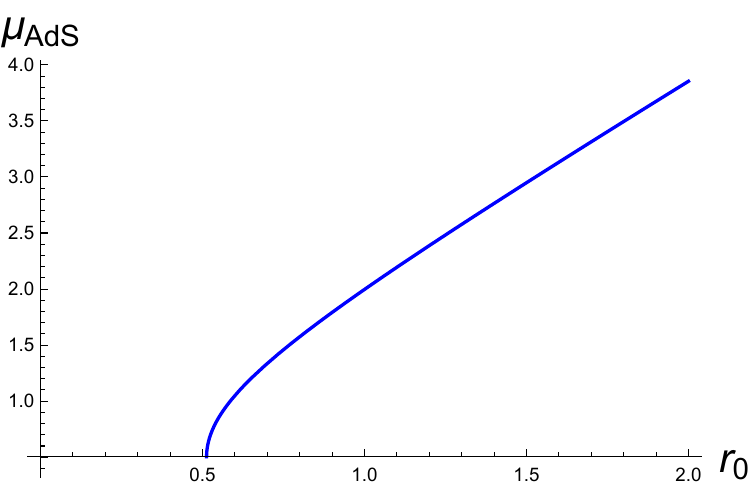}
 \label{fig:diagvbADS}
  \caption{The dependence of $\mu_{AdS}$ on $r_0$ for black hole with asymptotic hyperbolic boundary with  $q=0.1$.}
\end{figure}
It is interesting to compare these results  with the black hole with hyperbolic asymptotic boundary which can be obtained by analytical continuation $\ell \rightarrow i \ell$. In contrast with the   space of positive curvature where the chemical potential restores real-valued butterfly velocity one can observe  from Fig.\ref{fig:diagvbADS} the opposite effect (i.e. $v_b$ for some values of $r_0$ becomes complex-valued).
\section{Comparison with other models, related phenomena, and discussion}\label{sec:disc}
In the previous section, we have shown the degeneration of the butterfly velocity for the CFT at fixed local temperature $T$ propagating in the de Sitter universe with temperature $T_{dS}$. This butterfly velocity is inversely proportional to some parameter $\mu_{dS}=\frac{1}{2}\left(d-2+\sqrt{(d-2)^2-4 m_{scr}^2\ell^2}\right)$, where $m_{scr}$ is the function of $T$ and $\ell$. We find that for all values of $\ell$ and $T$ the butterfly velocity is complex-valued. On the other hand, we find that the chemical potential restores healthy positive butterfly velocity for a small enough $CFT$ temperature. In contrast to this for hyperbolic boundary, chemical potential creates some region of $r_0$ where butterfly velocity is complex-valued and vanishing for some finite $r_0$. Also, it is worth noticing, that the Euclidean $dS$ can be rewritten as an ordinary $d-$dimensional sphere in appropriate coordinate system. Thus the results obtained in this note can be considered as facts about CFT on the sphere.

$\,$

To understand the quantitative picture of this chaotic spreading disappearance, one has to make calculations in higher-dimensional CFT at a finite temperature in the de Sitter background. Instead of doing this, let us try to understand this result from the viewpoint of  different phenomena taking place in the de Sitter space and construct the qualitative picture  of its chaotic properties
\begin{itemize}

    \item As the non-equilibrium processes like the thermalization in holographic CFT (which are chaotic) should be accompanied somehow with the chaotic spreading, it is worth noticing the following result closely related to shock wave solutions. In de Sitter space, the black hole formation from the shock wave collisions is also sensitive to the values of Hubble parameter $H$ as it was shown in  \cite{Arefeva:2009oun,Arefeva:2009pxq,Arefeva:2011bcf}; for some range of $H$ and shock waves energies  the black hole formation is prohibited. From the viewpoint of the dual theory, this can be interpreted as the impossibility of thermalization and consequently partial disappearance of the chaotic properties\footnote{ The butterfly velocity and OTOC in the dual of dynamical $dS_{d+1}$  has been discussed in the context of $dS/CFT$ correspondence \cite{Aalsma:2020aib,Geng:2020kxh,Anninos:2018svg} and cosmology \cite{Choudhury:2020yaa, Haque:2020pmp,Bhargava:2020fhl}. }.
    \item  The butterfly velocity $v_B$  defines the effective lightcone constraining the spread of chaotic correlations  after the scrambling time  $\tau_*$. In general, the effect of chaos decrease or disappearance is expected because positively curved manifolds like $dS$  are ``antichaotic'' cousins of $AdS$ ( for example, in the sense of the geodesic behavior). However, the sharp transition to the degenerate complex-valued butterfly velocity indicates the presence of strong quantum effects.  The process inside the effective lightcone defined by $v_B$  can be thought of as the operator spreading effect resembling the branching diffusion \cite{Roberts:2018mnp}, which is influenced by the effect of dilution/concentration by curved spacetime.
        \item In some sense, a very similar effect that resembles the situation with the butterfly velocity is the Schwinger effect when the particles in de Sitter are created from the external electric field. In this process, particles and antiparticles exhibit the competing influence of electric field and the geometric properties of dS. The presence of different scales (de Sitter scale, particle mass, and electric field) sets different regimes of electric current.
        \item The analogy with the Schwinger effect and setting down different regimes of ``conductivity'' and current is supported by the proportionality of charge and thermal diffusion constant to a butterfly velocity established in \cite{Blake:2016wvh,Blake:2017qgd} for holographic field theories.

\item The example of how the chemical potential enhances particle production in cosmological setup is given in \cite{Bodas:2020yho,Sou:2021juh}. In general, the chemical potential allows the production of large mass particles, which should carry some chaotic imprint if being observed in the experimental setup.

\end{itemize}
It would be interesting to extend these results on the case when the boundary manifold is de Sitter and not the product of $dS$ and ${\mathbb{R}}$ as in our setting. Also, it is interesting to understand what happens when de Sitter is dynamical, i.e. to embed pole-skipping setup in $dS/CFT$ correspondence (we address this issue in forthcoming paper).

\section*{Acknowledgements}
This work is supported by the Russian Science Foundation under grant 24-72-10061; \href{https://rscf.ru/project/24-72-10061/}{ https://rscf.ru/project/24-72-10061/}. I would like to thank Viktor Jahnke, Juan Pedraza, and Saso Grozdanov for their comments on  this paper and correspondence.

\end{document}